\documentclass{optica-article}

\journal{opticajournal} 

\articletype{Research Article}

\usepackage{fancyhdr}
\usepackage{lineno}
\usepackage{siunitx}
\usepackage{layouts}
\usepackage{booktabs}
\usepackage{comment}
\usepackage{tikz}
\usepackage[symbol]{footmisc}
\usepackage{newtxtext,newtxmath}
\usepackage{xcolor}
\usepackage[utf8]{inputenc} 
\usepackage{amsmath}       

\DeclareRobustCommand\sampleline[1]{%
  \tikz\draw[#1, line width=1.8pt] (0,0) (0,\the\dimexpr\fontdimen22\textfont2\relax)
  -- (1.9em,\the\dimexpr\fontdimen22\textfont2\relax);%
}

\definecolor{Brown}{HTML}{8B4513}
\definecolor{Orange}{HTML}{F4A460}

\begin{document}
\title{Design and assembly of a cavity microscope with high numerical aperture for quantum simulations}

\author{G.S. Bolognini\authormark{1, *}, Z. Xue \authormark{1}, M. A. Eichenberger\authormark{1}, N. Sauerwein\authormark{1}, F. Orsi\authormark{1}, E. Fedotova\authormark{1}, R.P. Bhatt\authormark{1},  and J.P. Brantut\authormark{1}}

\address{\authormark{1}Institute of Physics and Center for Quantum Science and Engineering,
Ecole Polytechnique Fédérale de Lausanne (EPFL), CH-1015 Lausanne, Switzerland }

\email{\authormark{*}gaia.bolognini@epfl.ch} 

\begin{abstract*} 
We present the design and assembly of a cavity microscope for quantum simulations with ultracold atoms. The system integrates a high-finesse optical cavity with a pair of high-numerical aperture lenses sharing a common optical axis, enabling simultaneous operation with light close-to-atomic resonance. The system keeps the advantages of a rigid, single-block structure holding the lenses and cavity together, and improves over existing designs by using most of the solid angle left free by the cavity mode for imaging and atomic manipulation purposes. The cavity has a length of \SI{19.786}{\milli\meter}, a finesse of \SI{2.35}{\times 10^4} and operates \SI{214}{\micro\meter} away from the concentric limit, deep in the strong coupling regime. The two lenses offer a numerical aperture of $0.52$ each and maximal optical access in all directions transverse to the cavity axis, compatible with applications in quantum-gas microscopes, micro-tweezer arrays or few-fermions systems, as well as future cavity-assisted quantum simulation protocols demanding sub-cavity-mode control of the atom-cavity coupling. 
\end{abstract*}

\section{Introduction}
Cavity quantum electrodynamics (cQED) methods play an increasing role in neutral-atom-based quantum information and quantum simulation platforms \cite{Kimble_2008, Reiserer_2015, Mivehvar_2021}. In the context of quantum gases, upon reaching the strong collective coupling regime, the cavity photons mediate long-range interactions between atoms, allowing for the exploration of a wide range of interacting systems \cite{Varun_Ben_Lev_2018, Farokh_2019} and phases of matter \cite{Baumann_2010, Landini_2018, Leonard_2010, Muniz:2020aa, Phatthamon_2022, helson:2023aa, zhang:2021tr, Young_2024}. There, the atom-cavity system is driven by a pump laser tuned close to an atomic transition, yielding an interaction mediated by virtual cavity photon exchanges between atoms. This interaction inherits its spatial profile from the interference between the cavity and pump laser modes, which is typically of infinite range for standing or plane waves. Although the cavity mode shapes are determined by the resonator geometry, the shape of the pump beam can, in principle, be tuned using beam-shaping techniques, offering the potential for a wide variety of cavity-induced interactions, including arbitrary range and shapes as proposed in \cite{Bonifacio_2024}. The integration of cavity QED techniques with large aperture optics, similar to the capabilities provided by quantum gas microscopes \cite{Zupancic_16, Weitenberg_2011}, would provide the optical resolution required to control interactions at the microscopic level. Furthermore, the addition of large aperture optics could be employed for high resolution fluorescence imaging, enabling high-fidelity, parallel projective measurements of individual atoms \cite{Bergschneider_2018, gross:2021aa} as well as the measurement of complex observables such as high-order correlations \cite{Endres:2011aa, hilker:2017aa, rispoli:2019aa}, which ideally complements the continuous, quantum-limited readout permitted in cavity QED. 

The combination of an optical resonator with high-resolution imaging has expanded rapidly over the last years, such as in arrays of single-atoms in optical tweezer traps \cite{dordevic:2021aa,urunuela:2022aa,Stamper_Kurn_2022,liu:2023aa,Zhang_2024}. However, in quantum gases experiments, the large solid angle required for such imaging systems presents challenges due to the additional need for optical lattices, laser cooling or pumping beams in all directions. To minimize these constraints, the integration of an optical resonator with high-resolution imaging should be as compact as possible. Since the mode of a Fabry-Perot cavity occupies a negligible numerical aperture, an efficient approach is to align both systems along the same optical axis, as demonstrated in recent experiments \cite{wipfli2023integrationhighfinessecryogenic,orsi_2024}. Building upon those, here we present an improved cavity microscope system for quantum simulations with quantum gases. Compared with previous instances, our system enables to operate the cavity and the high-aperture optics simultaneously and at the same near-atomic-resonance wavelength. For example, it would make it possible to operate a quantum gas microscope inside the cavity, enabling homogeneous light-matter coupling for all sites in the focal plane of the microscope. It would also allow for arbitrary near-resonant beam geometries, intersecting the cavity axis at any angle within the aperture and in any plane containing the cavity axis. This system enables quantum simulation of fully-connected Fermi gas models with access to the tools of single-atom fluorescence imaging.\\
The paper is structured as follows: in section \ref{sec:Design_and_assembly}, we present an overview of the device and describe its design, assembly and alignment procedure. Section \ref{sec:performance} presents the overall performance of the system, with the characterization of both the imaging and the cavity properties of the device. We conclude by discussing some of the limitations of our approach and the open perspectives for quantum simulation. \\

\section{Design \& Assembly}\label{sec:Design_and_assembly}
\begin{figure}[b]
\centering\includegraphics[width=13cm]{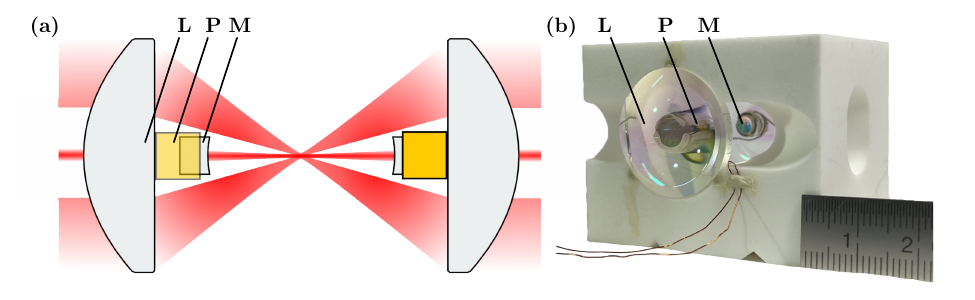}
\caption{Outline of the cavity microscope. (a) Schematics of the optical system. Two mirrors ('M') form a near-concentric optical resonator, confining the cavity-field (central red beam). The mirrors are attached to tube piezoelectric actuators ('P') which are glued to a pair of high numerical aperture (NA) lenses ('L'). The lenses focus at the cavity mode center, enabling high NA imaging on the atomic cloud and flexible beam coupling at various angles (side red beam). (b) Picture of the finalized device, where the two mirror-lens assemblies are glued to a monolithic macor holder which ensures the long-term stability of the optical resonator. } \label{fig:picture}
\end{figure}

Our system builds upon the cavity microscope concept demonstrated in \cite{orsi_2024}, improving on a number of aspects. Similar to \cite{orsi_2024}, it comprises a near-concentric Fabry-Perot cavity and a pair of aspherical lenses, aligned on a common optical axis. Fig.\ref{fig:picture}a presents an overview of the optical system, comprising a pair of mirror-lens assemblies. The cavity mode occupies only the central part of the setup, leaving a large portion of the solid angle available for imaging through the lenses. The small dimensions of the mirrors compared to their curvature avoid the steep surfaces that are challenging for polishing and coating \cite{Nelson_2019}. A tube piezoelectric actuator, used to tune dynamically the cavity length during operation, binds together the mirror and the lens in each of the assemblies. Fig.\ref{fig:picture}b shows a photograph of the fully assembled system with its holder, which acts as a spacer for the cavity, ensuring proper alignment of the optical elements and favoring the resilience of the setup against mechanical vibrations. The optical elements are fully glued in place, without any moving parts, ensuring long-term stability and limiting the risks of misalignment during experimental operation, crucial factors for a near-concentric cavity that is close to the instability point and highly sensitive to lateral drifts of the mirrors \cite{siegman1986lasers}. In this section, we describe the design and assembly of this system, both for the cavity-lens ensembles and for the final assembly including the holder.

\subsection{Design}
\begin{figure}[b]
\centering\includegraphics[width=13cm]{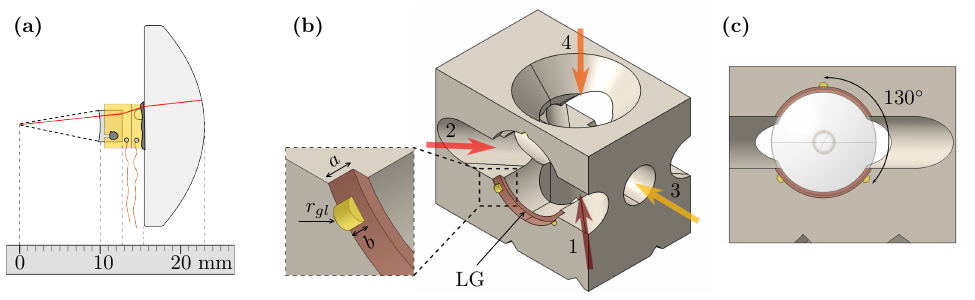}
\caption{Optical and mechanical design of the mirror-lens assembly and the cavity holder. (a) The three independent components of the mirror-lens assembly are glued together (glue spots shown in dark gray). The red line indicates the aperture limit for in and out coupling into the cavity, limited by the inner diameter of the piezoelectric actuators. The dashed line indicates the stability region of the bare cavity, dictated by the mirror radius of curvature. Thin electrical wires (in orange) connect the two piezo electrodes using conductive glue, shown in light gray. The piezoelectric tube includes a small venting hole at the location of the contact with the flat surface of the lens, ensuring proper outgassing during vacuum pumping. (b) The macor holder enables the coupling of multiple beams required for cold atoms experiment setup (exemplified by arrows 1-4). It ensures precise assembly of the cavity microscope, where the lenses are inserted into the lens groove (\SI{25.6}{\milli\meter} diameter and $a =$ \SI{2.5}{\milli\meter} depth) indicated as 'LG' and colored in brown and glued to the holder via the glue holes ($r_{gl} =$ \SI{1}{\milli\meter} radius and $b = $ \SI{1.5}{\milli\meter} depth), yellow colored and shown in the inset. (c) Three glue holes are arranged to evenly distribute the forces during the glue curing process. } \label{fig:assembly}
\end{figure}

The system includes a pair of mirror-lens assemblies fixed in place by the holder. Fig. \ref{fig:assembly}a shows an illustration of one of the cavity-lens elements, including a $D= \SI{4}{\milli \metre}$ diameter spherical cavity mirror and a standard, one inch diameter diffraction limited aspheric lens. These components are bonded by a $\SI{4.5}{\milli\meter}$ inner ($\SI{5.5}{\milli\meter}$ outer) diameter piezoelectric tube, employed for stabilizing the cavity length. The radius of curvature of the mirrors is $r = \SI{10}{\milli \metre}$, and the length of the piezoelectric tube (\SI{5}{\milli\meter}) allows for the lens focus to be located very close to the center of curvature of the mirrors, compatible with the cavity operating close to the concentric limit. Thin wires supplying the voltage for the piezo tube also run on the surface of the lens, without giving a significant degradation of the lens performance (see section \ref{subsec:imaging} for details). As shown in Fig.\ref{fig:assembly}a, the piezo tube effectively limits the aperture available for in and out coupling of light from the cavity, and determines the tolerance for misalignment of the cavity mode with respect to the optical axis. \\
The cavity lenses are attached to a Macor holder, a ceramic that combines machinability, stiffness and vacuum compatibility. A schematic with technical details is shown in Fig.\ref{fig:assembly}b. The holder cut-outs and holes are specifically designed to facilitate the cavity microscope assembly while providing the necessary optical access for atomic sample preparation and manipulation. To minimize in-vacuum complexity, the two mirror-lens elements are directly glued in place to the holder, fitting in the appositely designed lens grooves (LG in Fig.\ref{fig:assembly}b). To ensure that the glue does not shift the optical elements during its curing process, special cut-outs, with $\SI{1}{\milli\meter}$ radius and $\SI{1.5}{\milli\meter}$ depth, are designed to have a balanced application of the glue and to compensate the lateral forces upon curing, as shown in the inset of Fig.\ref{fig:assembly}b and in Fig.\ref{fig:assembly}c. Additionally, the holder features through holes for optical access at $50^{\circ}$ and $90^{\circ}$ of the cavity optical axis in the horizontal plane and along the vertical direction, providing the necessary optical access for laser cooling, trapping and imaging, and for the injection of the atomic beam, as shown in Fig.\ref{fig:assembly}b. Moreover, it shields the mirrors from the atomic beam contamination, blocking the line of sight between the mirror surfaces and the atomic source. \\
The holder geometry and the stiffness of Macor efficiently reduce the vibration sensitivity of the system, mainly affected by the motion of the piezos during cavity operation. The first mechanical resonance was experimentally observed on a test device at $\SI{67}{\kilo\hertz}$, which makes the use of further vibration cancellation methods for stable operation unnecessary. A finite element simulation identified this resonance as the mechanical mode involving mirror movement within the piezoelectric actuators. 

\subsection{Assembly}\label{sec:assembly}
\begin{figure}[htbp]
\centering\includegraphics[width=\textwidth]{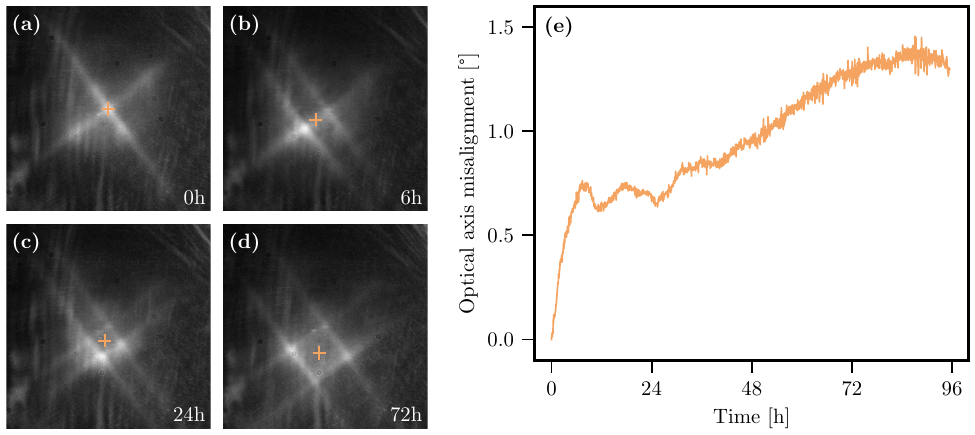}
\caption{Cavity misalignment tracking during the glue curing process. (a)-(d) Observed cavity transmission pattern at 0, 6, 24, 72 hours after the glue curing starts, showing a misalignment of the coupling to the $\text{TEM}_{00}$ mode. The orange crosses indicate the center of the cavity mode at different elapsed time. (e) Cavity axis misalignment with respect to the probe beam as a function of time elapsed over the 96 hours observation period.}\label{fig:misalignment}
\end{figure}

The first step is the assembly of the mirror-lens elements. We first attach electrical cables to the piezoelectric actuator using conductive UHV-compatible glue (EPO-TEK H20E). The cavity mirror is then inserted in the piezo tube and it is glued at the end of it through three symmetrically arranged holes using room-temperature curing epoxy (Masterbond Epoxy EP21TCHT-1), which is also employed in all subsequent gluing steps. The glue quantity is carefully balanced among the three points to avoid preferential strain directions, thereby minimizing externally induced birefringence which could arise in the mirrors. The free-end side of piezoelectric tube is then glued on the flat surface of the lens, where an additional hole on the side of the tube allows for outgassing during vacuum pumping. After room temperature curing overnight, the full cavity-lens assembly is baked at \SI{80}{\celsius} for \num{2} hours in a vacuum-oven, to  ensure a strong bond despite the small amount of glue employed. \\
The crucial point of this step is ensuring the alignment of the mirror with respect to the lens, such that the two share the optical axis. The centering of the mirror with respect to the piezo is checked before gluing using a binocular microscope, limiting misalignment to under \SI{100}{\micro\meter}. A custom-made Teflon centering tool is then used during the gluing of the piezo to the lens, ensuring an alignment accuracy of \SI{50}{\micro\meter} by construction.\\
The two mirror-lens assemblies are then aligned to form a near-concentric cavity, with the mode centered on the common optical axis of the lenses and cavity. We first perform a coarse alignment, using a pair of 5-axis translation stages and a probe laser with a continuously scanned frequency. The cavity transmission is monitored on a camera and on a photodiode, from which we measure the transverse mode spacing to evaluate the distance to the concentric limit. The coarse alignment leads to a cavity about \num{0.5} to \SI{1}{\milli \metre} away from concentricity. We then iteratively increase the cavity length while maintaining the alignment on the probe beam down to \SI{0.2}{\milli \metre} from the concentric limit. \\
Finally, the lenses are glued to the Macor holder with great care to ensure stability during the gluing and curing process. To address this, the lenses are attached sequentially, to compensate eventual misalignment of the first gluing step, and the glue is evenly spread on the three pads shown in Fig.\ref{fig:assembly}b and c while minimizing its volume to reduce mechanical stress released during the curing of the glue. Further details on the gluing procedure can be found in \cite{Bolognini_MasterThesis, Xue_Master_Thesis}. During curing, the alignment tools are left in place undisturbed and the cavity alignment is continuously monitored by checking the transmission of the probe beam via the camera. The laser frequency is swept over one free spectral range, allowing multiple transverse modes to couple to the cavity one after the other, within a timescale shorter than the camera integration time. The camera thus records the sum of the intensities of the coupled Hermite-Gaussian modes, weighted by the square of their overlap with the input beam. 
Examples of profiles recorded during the curing process are shown in Fig.\ref{fig:misalignment}a-d. Despite the initial optimal coupling to the $\text{TEM}_{00}$ the profile presents a cross-line pattern, Fig.\ref{fig:misalignment}a, which we attribute to astigmatism and imperfections of the incoupling beam. As the glue polymerizes, curing-induced mechanical stress displaces the lens–mirror assembly, reducing  coupling to the $\text{TEM}_{00}$ mode and increasing it to higher-order modes. The pattern evolves into a square-like shape as shown in Fig\ref{fig:misalignment}b-d, which has symmetry about the preferential directions emerging from the cavity astigmatism (see section \ref{sec:Cavity}). The pattern can be well reproduced in a simulation from the projection of an input Gaussian beam displaced with respect to the cavity center over Hermite-Gaussian modes.
The center of mass of the recorded intensity pattern thus measures the misalignment of the cavity optical axis relative to the probe as shown in figure \ref{fig:misalignment}e. In line with the expected curing process of the glue, we observe first a fast evolution followed by a gradual stabilization towards a fixed steady direction, corresponding to an overall drift of the cavity axis by \SI{1.3}{\degree}. This is about five times smaller than the maximally tolerated drift limited by the clear aperture of the setup (red beam in Fig.\ref{fig:assembly}a). After the curing time of the glue, we recovered the optimal coupling to the $\text{TEM}_{00}$ mode by realignment of the pilot beam. Interestingly, we observed that the cavity returned to almost $0^{\circ}$ misalignment after relaxation of the glued setup for several weeks.

\subsection{Robustness and vacuum compatibility}
After curing, the cavity microscope was tested for vacuum compatibility, by placing it in a test chamber pumped with a turbomolecular pump and equipped with a residual gas analyzer. Care was taken to avoid fast pressure variations at the start of the pumping process and at the venting, anticipating that the small venting holes in the piezoelectric tubes may lead to transient pressure differences affecting the alignment. We first evacuated the chamber at room temperature, until the partial pressure of both hydrogen and water fell below the \SI{}{10^{-9} \milli bar}, confirming vacuum compatibility of the ensemble. We then performed slow temperature ramps in vacuum, after which the cavity was taken out of the chamber and the alignment tested with the same procedure as for the characterization (described in section \ref{sec:Cavity}). For ramps up to \SI{40}{\degree C} we did not observe any measurable drift of the cavity, nor variations in the finesse. When ramping to \SI{50}{\degree C} we observed a cavity axis misalignment up to \SI{1.7}{\degree}, which we attribute to plastic deformation of the glue upon thermal expansion. The cavity finesse remained unchanged. These performance suggest that baking of the final quantum gas experimental system slightly above room temperature should be feasible, similar to the baking procedure used in previous experiments \cite{sauerwein2023engineering} which led to atom lifetimes at the minute scale. \\

\section{Characterization}\label{sec:performance}
\subsection{Imaging} \label{subsec:imaging}
The performance of the lens was characterized before and after the assembly with the mirror and piezos, configurations to which we will refer as \textit{non obscured} and \textit{obscured} respectively. In our assembly, we use the Thorlabs AL2520H aspheric lens, with focal distance \SI{20}{\milli\meter}, working distance \SI{15.7}{\milli\meter} and diffraction limited at \SI{780}{\nano\meter}. The clear aperture is \SI{21.3}{\milli\meter}, resulting in a numerical aperture of \num{0.52}, common for cold atoms experiments \cite{Sortais:2007aa, Tey:2008aa,Nogrette:2014aa}. Previous experiments suggest that the imaging performance and trapping capabilities are not affected by obscuring the central part of the lens \cite{lorenz:2021aa, wipfli2023integrationhighfinessecryogenic, su:2025aa}. In our case, the mirror obscures a disk representing \num{0.37} radial fraction of the clear aperture, which translates to a transmission of around \num{86}\% of the obstruction-free case for a uniformly illuminated lens. It is also expected to filter the low spatial frequency components reducing their contrast, but to leave the resolution and the numerical aperture essentially unaffected \cite{Everhart1959DiffractionPP}, since these depend primarily on the imaging system’s high-frequency components. \\

\begin{figure}[htbp]
\centering\includegraphics[width=1\textwidth]{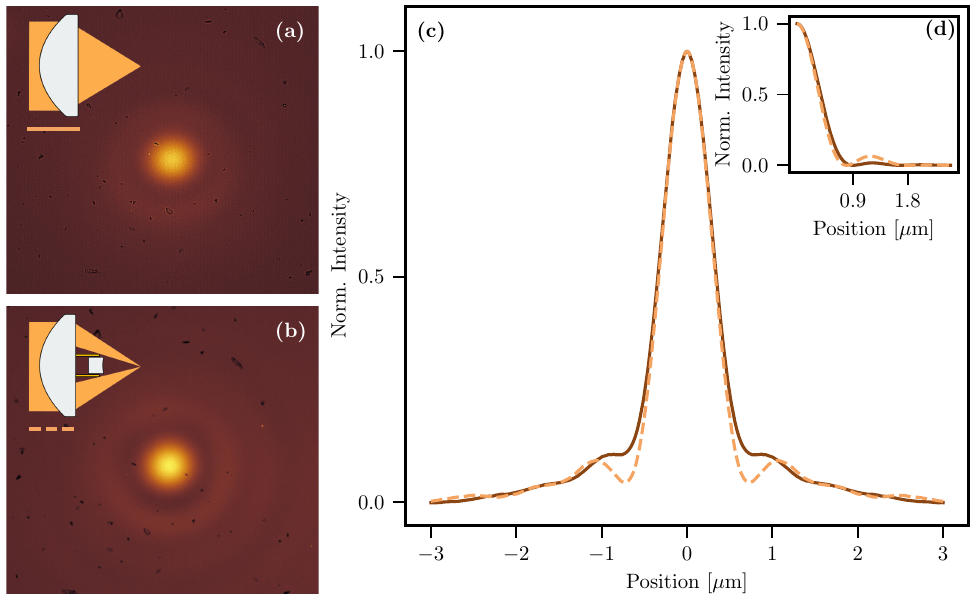}
\caption{Imaging properties of the cavity microscope. (a)-(b) Point spread function of the imaging system as obtained on a CCD camera for the bare lens and obstructed lens, schematically represented in the upper-left side. (c) Comparison of the radial averaged cut of the pictures for the bare lens (\textcolor{Brown}{\sampleline{}}) and the obstructed lens (\textcolor{Orange}{\sampleline{dashed}}). (d) Theoretically simulated curve for the same two cases, where the first minimum indicates the diffraction limited resolution of the imaging system. }\label{figure:4}
\end{figure}

The lens performance was tested by imaging a point source in the focal plane of the lens. The point source was produced from a sub-wavelength-diameter pinhole and placed in the far field behind the lens. The focal plane was then imaged using a commercial microscope objective onto a CCD camera, producing an image of the Point Spread Function (PSF) of the lens. Fig.\ref{figure:4}a and b show examples of the PSF observed before and after attaching the mirror and the piezoelectric tube at the center for a source wavelength of \SI{780}{\nano\meter}. Fitting a Gaussian to the central feature of the PSF yields a $1/e$-radius of \SI{0.31(2)}{\micro\meter} and \SI{0.30(3)}{\micro\meter}, before and after obscuring, respectively, confirming the expectation that the tube at the center does not alter the resolution. The main effect of the obstruction is the pronounced ringing effect at the edges of the Airy disk, as can be seen in Fig.\ref{figure:4}c. This is the expected effect of the spatial frequency filtering, in qualitative agreement with the predictions from diffraction, as shown in Fig.\ref{figure:4}d.\\
Anticipating the use for cold atom experiments, the lenses were AR coated for multiple wavelengths to suit different use cases. Specifically, we tested the lens at \SI{671}{\nano\meter}, resonant to the atomic transition of lithium atoms, for which very similar performance were obtained in the both obscured and non-obscured cases.

\subsection{Cavity}\label{sec:Cavity}

\begin{figure}[htbp]
\centering\includegraphics[width=1\textwidth]{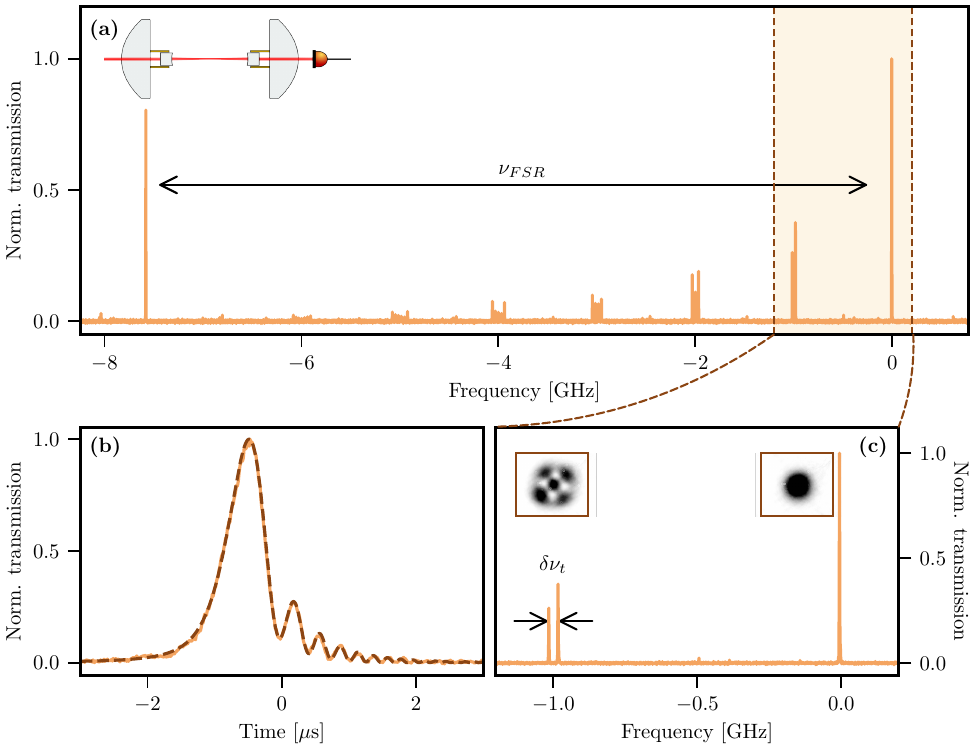}
\caption{Cavity spectrum and characterization. (a) Cavity transmission spectroscopy when scanning the laser frequency over more than one free spectral range. The distance between two consecutive longitudinal $\text{TEM}_{00}$ modes of the cavity is highlighted with the arrow ($\nu_{\text{FSR}}$). The schematics of the experimental apparatus is shown on the upper-left side. (b) Cavity ringdown measurement (\textcolor{Orange}{\sampleline{}}) and corresponding fitting function (\textcolor{Brown}{\sampleline{dashed}}) to extract the linewidth of the cavity $\kappa$ and its finesse. (c) Zoom-in on $\text{TEM}_{00}$ and $\text{TEM}_{20}$- $\text{TEM}_{02}$ of the upper panel, accentuating their non-degeneracy ($\delta \nu_{\text{t}}$). The brown squares show a picture of the spatial distribution of the two modes as observed on a CCD camera.} \label{figure:5}
\end{figure}

The cavity mirrors have $r = \SI{10}{\milli \metre}$ radius-of-curvature substrates and are super-polished to a nominal root-mean-square surface roughness of below \SI{0.15}{\nano \metre} (Perkins Precision). They are specified to reach high finesse at \SI{671}{\nano \metre} and \SI{1342}{\nano \metre} with design transmission of \SI{15(5)}{ppm} and \SI{25(5)}{ppm} respectively (FiveNine Optics).\\
After the final assembly of the cavity microscope, we perform transmission spectroscopy to characterize the cavity \cite{Bachor_2004}. Fig.\ref{figure:5}a presents the transmission spectrum spanning one free spectral range (FSR). For a well-aligned probe laser, the coupling to odd transverse modes vanishes, and we only observe even mode profiles in the transmission, due to imperfect beam profile matching between the probe and the fundamental cavity mode as shown in figure \ref{figure:5}c. Combining the spectroscopic data with the mode profile measurement, we measure the free spectral range of the cavity $\nu_{FSR}= \SI{7.581(6)}{\giga\hertz}$ and the transverse mode spacing $\nu_t = \nu_{FSR} - \SI{500.1(1)}{\mega \hertz}$. From these two parameters and the knowledge on the mirrors radius of curvature stated by the manufacturer, we precisely determine the cavity length \SI{19.786(14)}{\milli\meter}, with a corresponding distance from concentricity of \SI{214.0(1)}{\micro\meter}, and the fundamental mode waist $w$ of \SI{14.82(1)}{\micro\meter}. We also observe a frequency splitting between the TEM$_{02}$ and TEM$_{20}$ equal to $\delta\nu_{t} =$ \SI{32.8(1)}{\mega\hertz} (in figure \ref{figure:5}c), which we attribute to astigmatism, amounting to a difference in radius of curvature of \SI{7}{\micro\meter} between two orthogonal directions. \\
To measure the finesse of the cavity, we sweep the frequency of the probe laser through the fundamental mode resonance at finite speed and observe ringing in the transmission \cite{Poirson_97}. An example measurement is shown in Fig.\ref{figure:5}b. We fit this pattern using the solution of the time-dependent equation \cite{walls2008quantum}
\begin{equation} \label{eq::ring_diff_eq}
        \frac{\text{d}\alpha(t)}{\text{d}t}=-\frac{\kappa}{2}\alpha(t)+\sqrt{\kappa}\alpha_{\text{in}}(t)
\end{equation}
with $\alpha_{\text{in}}(t)$ a frequency chirp, and $\kappa$ the cavity photon lifetime. The resulting fit is shown in dashed line in figure \ref{figure:5}b, showing excellent agreement, yielding $\kappa=2\pi\times \SI{323(8)}{\kilo\hertz}$ and a finesse $\mathcal{F}=$\SI{2.35(6)}{\times 10^4}. Performing the same procedure with a probe laser at \SI{1342}{\nano \metre}, we obtain a finesse of \SI{3.42(7)}{\times 10^4} despite the larger transmission specification. For this set of mirrors, we estimate a loss level of \SI{119(4)}{ppm} at \SI{671}{\nano \metre}, from the transmission design parameter \cite{Hood_2001} and attributing the anomalies to losses. A better performance was observed for mirrors from the same coating batch but deposited on standard superpolished substrates with high radius-of-curvature, pointing at the need for better polishing technologies for centimeter-scale radius of curvature substrates \cite{Jin_22}.\\
Based on the mode waist estimate and the finesse, we determine the cooperativity for near-resonant light $\frac{24\mathcal{F}}{\pi k^2w^2}= 9.3(2)$ \cite{Tanji_Suzuki_2011} (with $k$ the wave number of light), which is the fundamental figure of merit for most of quantum simulation, sensing and information processing applications \cite{H_J_Kimble_1998}. The properties of the cavity are summarized in table \ref{tab:Cavity_Properties}.

\begin{table}[h]
\centering
		\begin{tabular}{c c c}
			\toprule
			Wavelength & \SI{671}{\nano\meter} & \SI{1342}{\nano\meter}  \\
			\midrule
            Cavity length $L$ & \multicolumn{2}{c}{\SI{19.786\pm0.014}{\milli \meter}}  \\
			Free spectral range $\nu_{\text{FSR}}$ & \multicolumn{2}{c}{\SI{7.581\pm0.006}{\giga\hertz}}  \\
			Distance from concentricity $2r-L$ & \multicolumn{2}{c}{\SI{214.0\pm0.1}{\micro \meter}} \\
			Transverse mode spacing $\nu_{\text{t}}$ & \multicolumn{2}{c}{$\nu_{\text{FSR}}-\SI{500.1\pm0.1}{\mega\hertz}$} \\
			Linewidth $\kappa/2\pi$ & \SI{323\pm8}{\kilo\hertz} & \SI{222\pm4}{\kilo\hertz} \\
			Finesse $F$ & \SI{2.35\pm0.06}{\times 10^4} & \SI{3.42\pm0.07}{\times 10^4} \\
			Mode waist $w$ & \SI{14.82\pm0.01}{\micro \meter} & \SI{20.96\pm0.01}{\micro \meter} \\
			Cooperativity $\eta$ & \SI{9.3\pm0.2}{} & \\
			\bottomrule
		\end{tabular}
		\caption{Summary of cavity properties.}\label{tab:Cavity_Properties}
\end{table}

\section{Discussion}\label{sec:Discussion}

We have presented the fully integrated device comprising a pair of lenses and a cavity, operating with a common optical axis, enabling the use of the high-aperture optics of the imaging system to collect fluorescence and to project pump laser beams. Importantly, our cavity does not feature any measurable birefringence in the transmission spectrum, in contrast with ring cavities allowing for small beam waists \cite{Culver_2016, Zhang_2024}. It thus supports circularly polarized light driving the closed transitions and we do not expect polarization effects to reduce the cooperativity. 

After having tested the system as described above, we assembled and tested a second version of the cavity microscope using an identical procedure, reaching again $\approx$\SI{200}{\micro\meter} distance from concentricity, suggesting that this value is rather conservative with respect to the precision of our curing procedure. The main intrinsic limitation remains the finesse that can be reached, with mirrors showing consistently higher losses than the specified transmission of the coatings at \SI{671}{\nano\meter}. For our best set of optical mirrors, we observed a factor $\approx 3$ higher losses at \SI{671}{\nano \metre} than at \SI{1342}{\nano \metre} suggesting Rayleigh scattering as their origin. Our second prototype yields a finesse at \SI{671}{\nano \metre} of $\mathcal{F}$ = \SI{4.61(7)}{\times 10^4} and a cooperativity of \SI{18.3(3)}. For atoms with longer wavelength transitions, losses in the currently available superpolished substrates should allow conservatively for an increase of the finesse by at least a factor of three. Our approach could be combined with existing methods to push the cavity length closer to the instability point \cite{PhysRevA.98.063833, adam2024low}, reaching cooperativities of the order of $\sim150$, on par with state-of-the-art quantum information processing systems \cite{Mivehvar_2021, Kimble_2018}. \\
Our system will provide a decisive advantage for quantum simulations which integrate optical cavities with optical lattices. There, projecting both the optical lattice and the pump beams through the same lens system ensures passive phase-stability between the lattice and the pump-cavity interference pattern, essential for the realization of exotic phases in Fermi-Hubbard models involving topology \cite{Santiago_2017, Jaksch_2019} or dynamical gauge fields \cite{Kollath_2016, Colella_2019}. The pump beam angle can be continuously tuned relative to the cavity axis, enabling exploration of novel phases influenced by commensurate or incommensurate lattice effects \cite{Dogra_2016, Morigi_2013}. Furthermore, by using all the available pump-cavity angles to realize a speckle pattern for the pump geometry, the cavity-mediated interaction will feature a dense, continuous range of wavevectors involved in photon-exchanges. This eventually results in fully connected, randomly interacting Fermions - a setting of fundamental interest in the study of quantum many-body physics \cite{chowdhury:2022aa , uhrich2023cavity, baumgartner2024quantumsimulationsachdevyekitaevmodel}.

\begin{backmatter}
\bmsection{Acknowledgments}
We thank Claude Amendola and the staff of the mechanics workshop of EPFL for discussions and assistance. We thank  Jonas Faltinath for experimental help during the first stages of the experiment, Giulia del Pace for useful discussions and Léa Aurélie Dubois for careful reading of the manuscript. 
\bmsection{Funding}
We acknowledge funding from Swiss State Secretariat for Education, Research and Innovation (Grants No. MB22.00063 and UeM019-5.1) and from the Swiss National Science Foundation (Grant No. 200020E\_217124/1). 
\bmsection{Disclosures}
The authors declare no conflicts of interest.
\bmsection{Data Availability}
The data files corresponding to this article are available from the Zenodo repository \cite{Zenodo_folder}.

\end{backmatter}

\bibliography{Cavity_design_Submission_Review}
\end{document}